\documentclass[apj,numberedappendix,revtex4,twocolappendix]{emulateapj}
 \usepackage{natbib}
 \usepackage{subfigure}

\newcommand{\etal}{{\it et al.}}
\newcommand{\ie}{{\it i.e.}~}
\newcommand{\eg}{{\it e.g.},~}

\newcommand{\Rmnum}[1]{\footnotesize{\uppercase\expandafter{\romannumeral #1}}}

\usepackage[breaklinks,colorlinks,urlcolor=blue,citecolor=blue,linkcolor=blue]{hyperref} 

\hypersetup{
    colorlinks=true,       
     linkcolor=blue,          
     citecolor=blue,        
     filecolor=black,      
      urlcolor=blue           
}

\shorttitle{Filament-Filament Interaction}
\shortauthors{Zhu \etal}

\begin{document}
\ifx \doiurl    \undefined \def \doiurl#1{\href{http://dx.doi.org/#1}{\textsf{#1}}}\fi
\ifx \adsurl    \undefined \def \adsurl#1{\href{http://adsabs.harvard.edu/abs/#1}{\textsf{#1}}}\fi
\ifx \arxivurl  \undefined \def \arxivurl#1{\href{http://arxiv.org/abs/#1}{\textsf{#1}}}\fi

\title{\uppercase{Complex Flare Dynamics Initiated by a Filament-Filament Interaction}}

\author{Chunming Zhu\altaffilmark{1,2}, 
Rui Liu\altaffilmark{3,4}, 
David Alexander\altaffilmark{2}, 
Xudong Sun\altaffilmark{5}, 
R.T. James McAteer\altaffilmark{1}}
\affil{$^{1}$Department of Astronomy, New Mexico State University, NM 88003, USA; \href{mailto:czhu@nmsu.edu}{czhu@nmsu.edu}}
\affil{$^{2}$Department of Physics and Astronomy, Rice University, TX 77005, USA}
\affil{$^{3}$CAS Key Laboratory of Geospace Environment,
               Department of Geophysics and Planetary Sciences, 
               University of Science and Technology of China,
               Hefei 230026, China}
\affil{$^{4}$Collaborative Innovation Center of Astronautical Science and Technology, China}
\affil{$^{5}$W. W. Hansen Experimental Physics Laboratory, Stanford University, Stanford, CA 94305, USA}

\begin{abstract}

We report on an eruption involving a relatively rare filament-filament interaction on 2013 June 21, observed by SDO and STEREO-B.  The two filaments were separated in height with a `double-decker' configuration. The eruption of the lower filament began simultaneously with a descent of the upper filament resulting in a convergence and direct interaction of the two filaments. The interaction was accompanied by the heating of surrounding plasma and an apparent crossing of a loop-like structure through the upper filament. The subsequent coalescence of the filaments drove a bright front ahead of the erupting structures. The whole process was associated with a C3.0 flare followed immediately by an M2.9 flare. Shrinking loops and descending dark voids were observed during the M2.9 flare at different locations above a C-shaped flare arcade as part of the energy release, giving us a unique insight into the flare dynamics.
\end{abstract}


\section{\uppercase{Introduction}}
Solar filaments represent the observational manifestations of relatively cool and dense plasma suspended in the solar corona.  Their eruptions are often associated with solar flares and Coronal Mass Ejections (CMEs; see review by \citealp{forbes2000review}). Filaments always form in filament channels (\citealp{gaizauskas1997formation}; \citealp{gaizauskas1998filament}; \citealp{wang2007formation}), which are a particular type of magnetic structure located above and along the polarity inversion line (PIL) where the photospheric magnetic field changes polarity.

It has been reported that the interaction between filaments displaying the same sign of chirality at their neighboring \textit{endpoints} can result in them merging into a longer single filament \citep{schmieder2004magnetic,bone2009formation}. The new filament formed by the so-called ``head-to-tail'' linkage is thought to remain steady until further magnetic cancellation at the footpoints \citep{litvinenko2000magnetic, martens2001origin, devore2005solar},  which eventually results in its destabilization and eruption.

Collisions and interactions between the \textit{bodies} of filaments have also been studied both observationally and in numerical simulations. Based on a series of MHD simulations regarding the interaction of twisted flux tubes under convective zone conditions, \citet{Linton2001} proposed four types of fluxtube interaction: bounce, merge, slingshot and tunnel. \citet{su2007observation} reported the merger of two filaments where the sudden injection of mass from one to the other triggered an eruption. \citet{jiang2014interaction} studied the merger of two sinistral filaments that resulted in a significant heating of nearby plasma.  \citet{kumar2010evolution} reported the ``collision'' of the central segments of two filaments and a subsequent creation of newly formed stable filaments with their footpoints  exchanged, corresponding to the slingshot reconnection  as described by \citet{Linton2001}, \citet{Linton2006}, and \citet{Toeroek2011}. An M1.4 flare was also reported during this event, but not directly related to the filament interactions \citep{chandra2011homologous}. \citet{Jiang2013} investigated a similar event suggesting a partial slingshot reconnection between a small filament and a nearby larger and denser filament, although no flare was observed. \citet{alexander2006hard} studied the eruption of a kinking filament and identified a hard X-ray coronal source at the projected crossing of the writhing filament structure, which was regarded as a process similar to the interaction of two converging flux tubes (the two legs of the same filament in this case). However, detailed observations on the interactions between filaments are still extremely rare, and the relationship between the interaction and flare energy release remains unclear.               

In this study, we present an analysis of an interaction between two filaments and its associated flaring. This article is organized as follows: in Section 2, we describe the observations on the interaction and eruption of both filaments and the accompanying emissions. Our interpretation is presented in Section 3. The concluding remarks are given in Section 4.

\section{\uppercase{Observations}}
\subsection{Instruments and Data}
\begin{figure*}[ht!]
  \begin{center}
    \includegraphics[bb =68 608 708 824, clip, width=0.89\textwidth]{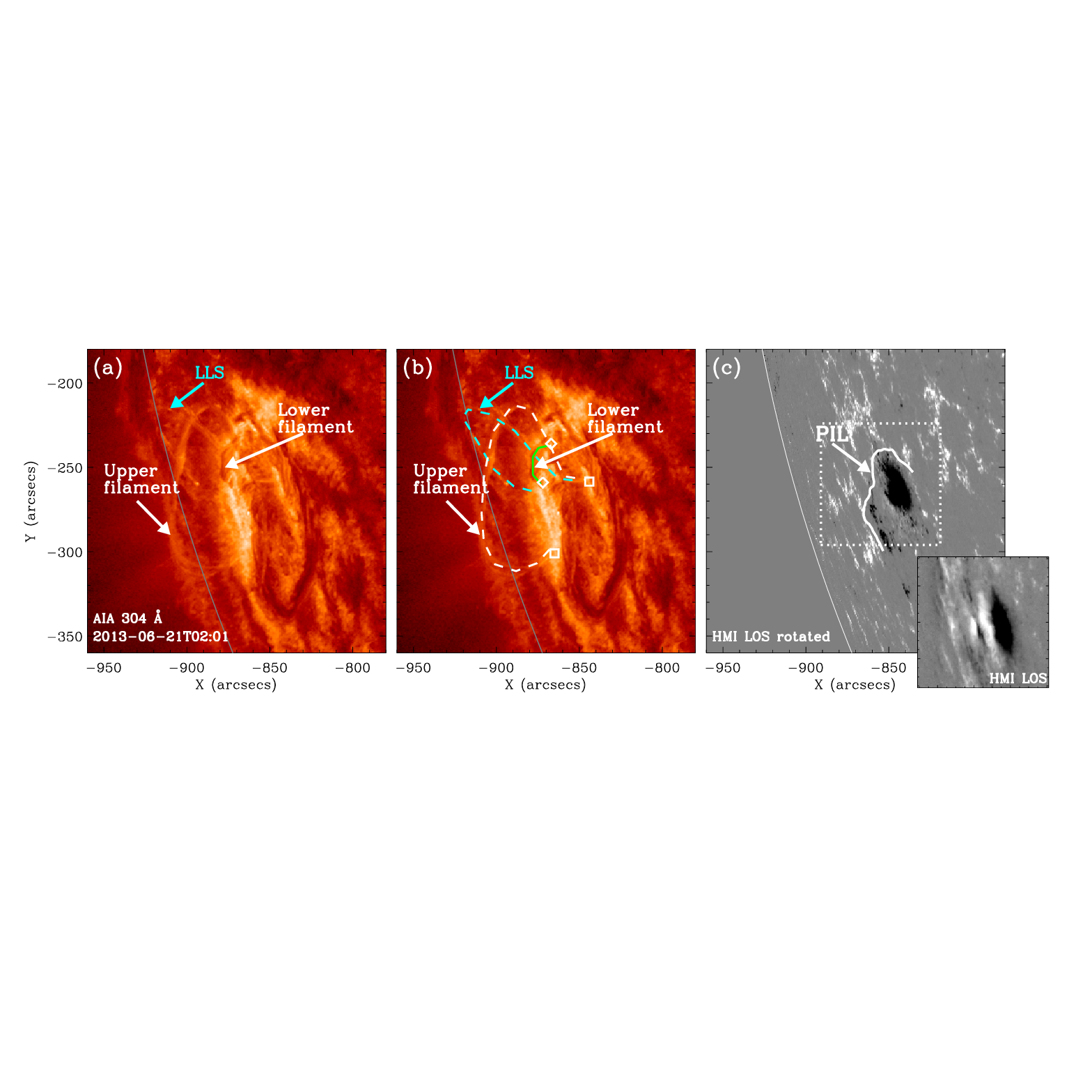}
    \includegraphics[bb = 40 472 457 729, clip, width=0.38\textwidth]{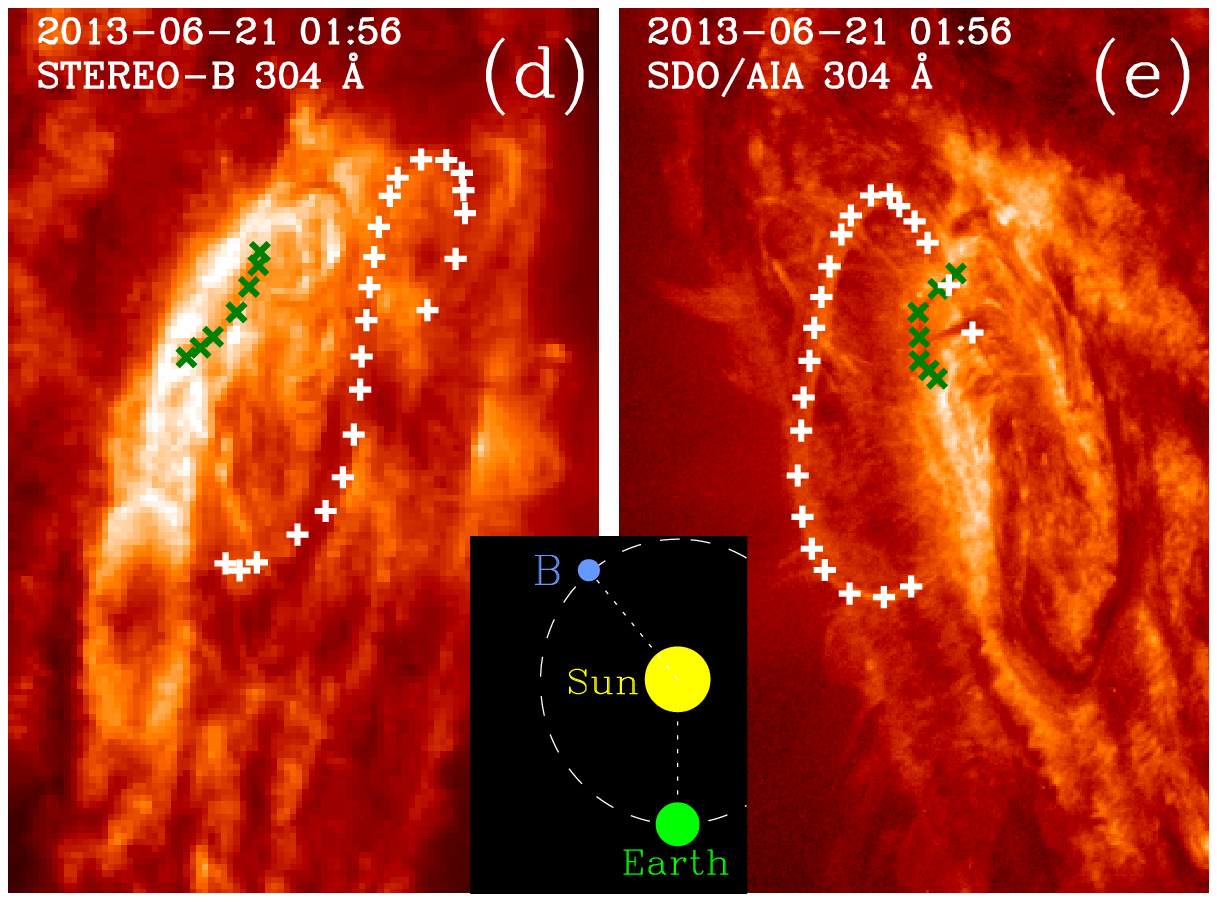}
    \includegraphics[bb = 64 160 301 689, clip,angle=90, width=0.52\textwidth]{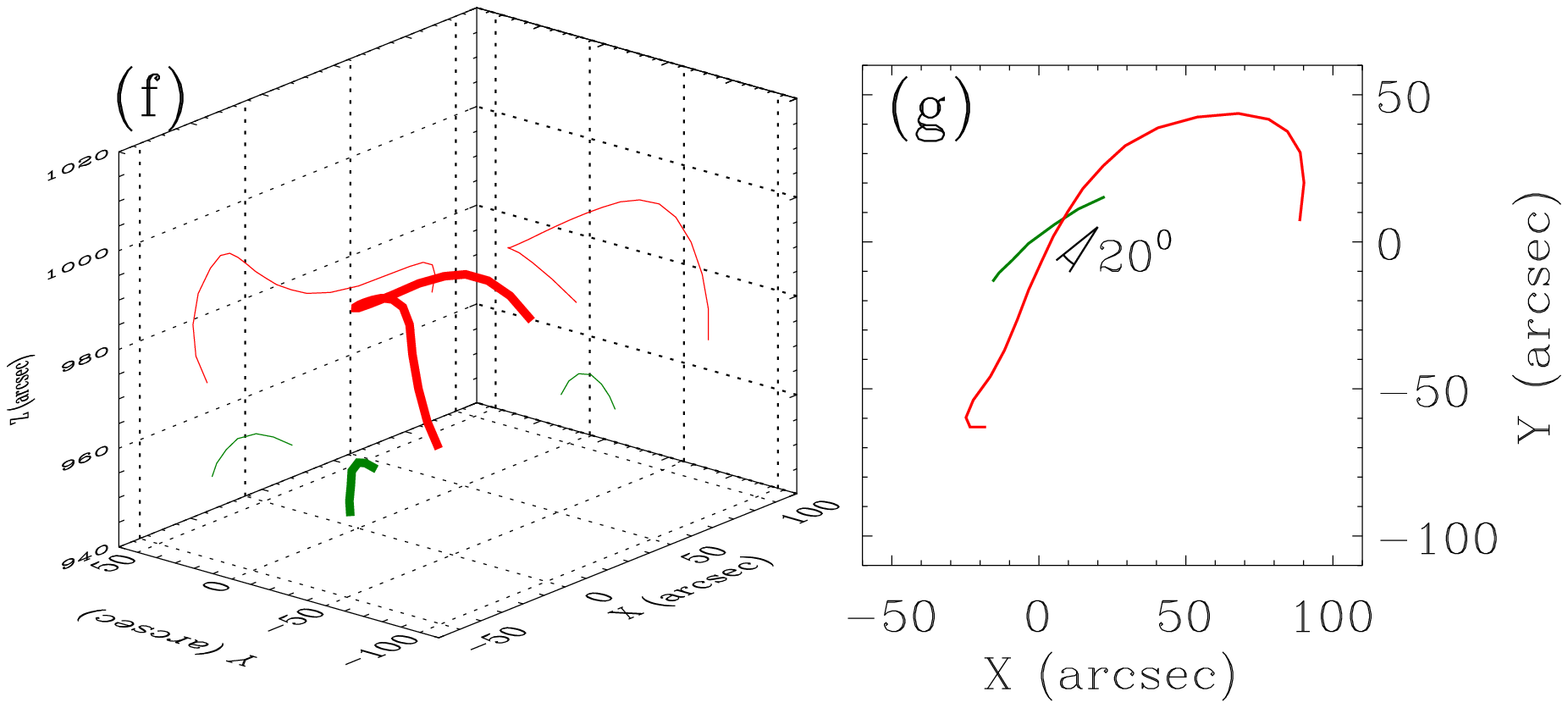}

    \caption[fig1]{Configurations of the lower and upper filaments before eruption. (a): 304~\AA ~image showing the filament structures. A loop-like structure (LLS) remained above the upper filament. (b): The curves denote the structures from (a). The apparent footpoints of the upper and the lower filaments are marked with symbols of squares and diamonds, respectively.  (c): A line-of-sight (LOS) HMI magnetogram from four days later, differentially rotated to the same time in (a). The thick curve indicates the location of the PIL in the active region.  A small panel displays the LOS fields in the dotted box at 02:01 UT. (d) and (e): Observations from both viewpoints of STEREO-B (left) and SDO (right), with their relative positions shown in the middle panel. The corresponding features, indicated by the plus and cross signs, were used for the 3D reconstruction of the upper and the lower filaments. (f): 3D configuration of both filaments, with the middle of the lower filament rotated to the center of disk, and their projections on the X-Z and Y-Z planes. (g): The projections of both filaments on the X-Y plane, indicating a small crossing angle of around 20$^{\circ}$ between them.}
    \label{fig:fig1}
  \end{center}
\end{figure*}

The filaments under study were located in NOAA AR~11777 (E71S16, Figure~\ref{fig:fig1}), appearing near the eastern limb on 2013 June 21, as viewed from the perspective of the Solar Dynamics Observatory (SDO). The Atmospheric Imaging Assembly (AIA; \citealp{Lemen2012}) onboard SDO takes full-disk images of the Sun in seven extreme-ultraviolet (EUV) channels ($\log$T ranges 5.6$\sim$7.3) and three UV to visible channels ($\log$T ranges 3.7$\sim$5.0), with a pixel scale of 0.6$''$ pixel$^{-1}$  and a cadence of 12 seconds. The Helioseismic and Magnetic Imager (HMI; \citealp{Schou2012}) onboard SDO provides full-disk vector magnetograms (\citealp{hoeksema2014helioseismic}) with a pixel scale of 0.5$''$ pixel$^{-1}$ and  a cadence of 12 minutes. 

The evolution of the filaments was also observed near the western limb from the perspective of the Solar Terrestrial Relations Observatory Behind spacecraft (STEREO-B), with a separation angle of approximately 140$^{\circ}$ away from SDO. The \textit{Extreme-Ultraviolet Imager} (EUVI; \citealp{wuelser2004euvi}) on board the STEREO took images at four bandpasses centered at 171~\AA, 195~\AA, 284~\AA, and 304~\AA, with spatial resolution of 1.6$''$ pixel$^{-1}$. The STEREO/EUVI had an image cadence of 5 minutes in the 195~\AA ~filter and a 10 minute cadence at 304~\AA, for the data used in this study.

 \begin{figure*}[ht!]
  \begin{center}
   \includegraphics[bb = 52 261 663 409, clip, width=0.98\textwidth]{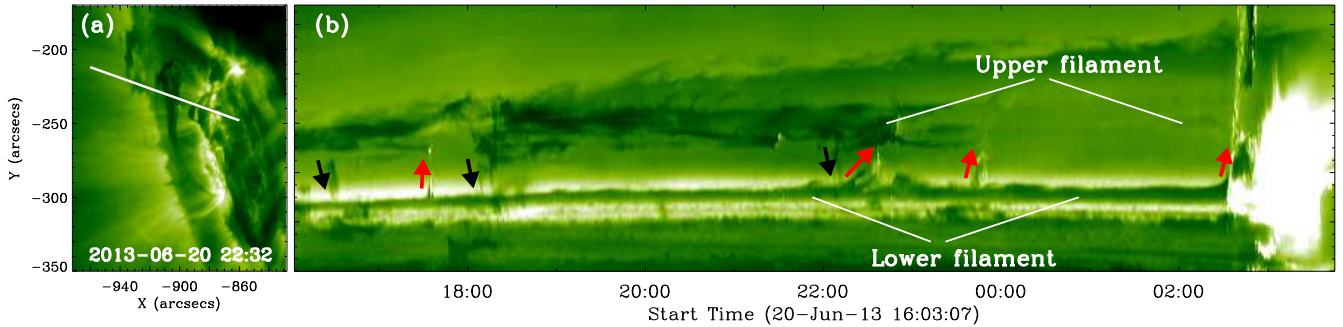}
      \caption[fig2]{Material motions between the lower and the upper filaments, observed by SDO/AIA 193~\AA. (a): A slit through two filaments was chosen for the stackplot in (b). (b): Material was observed to transfer from the lower filament to the upper one, marked by the red arrows. The black arrows indicate mass drainage. These motions can be clearly seen in Movie~2.}
    \label{fig:fig2}
  \end{center}
\end{figure*}
  
  \begin{figure*}[ht!]
  \begin{center}
      \includegraphics[bb =79 162 652 600, clip, width=0.95\textwidth]{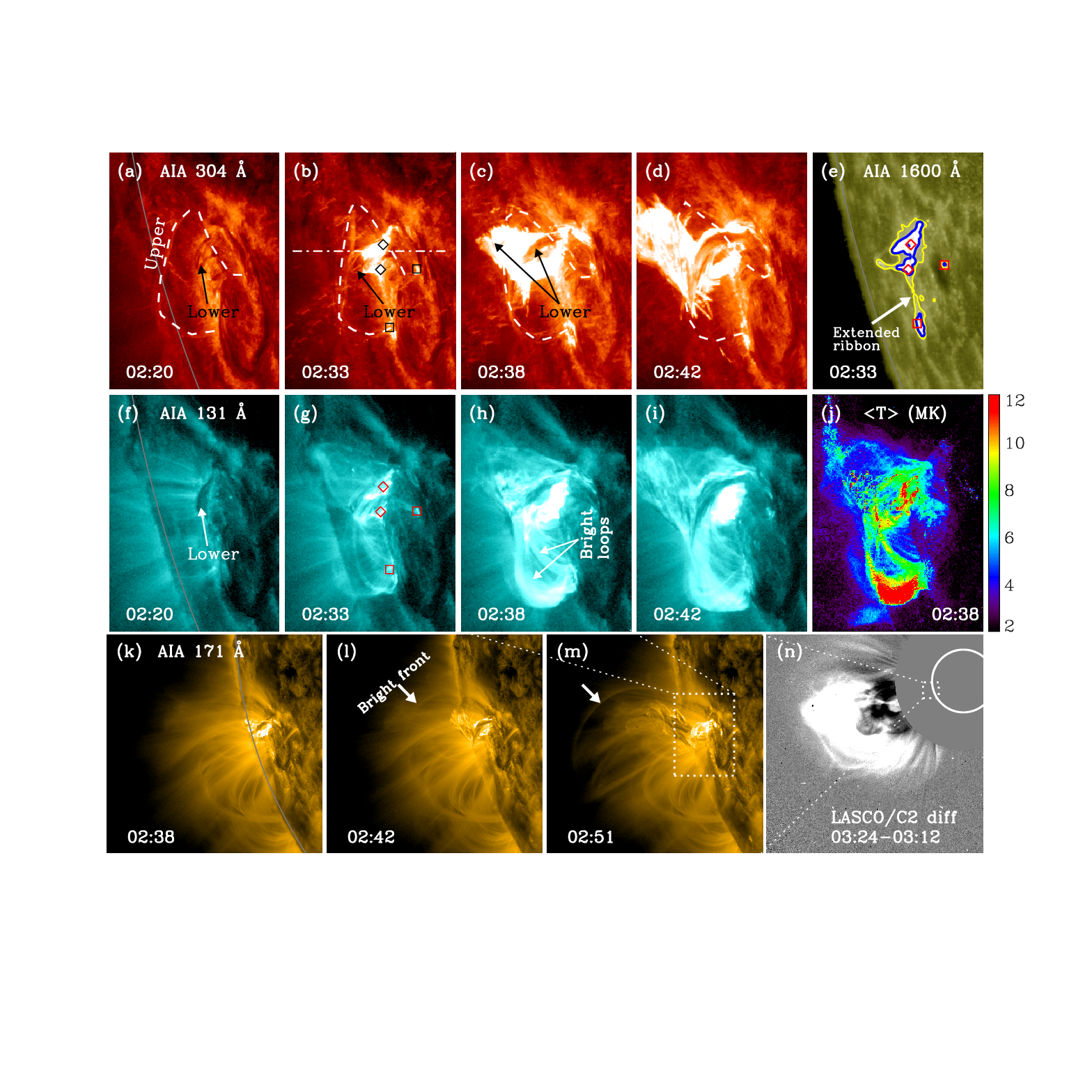}
    \caption[fig3]{Evolution of the interaction and eruption of the two filaments. A localized region of hot plasma appeared at their interaction. Panels (a)-(d): observations from AIA 304~\AA. (e): AIA 1600~\AA ~image at 02:33~UT, with its contours ~at 400 DN s$^{-1}$ (in yellow) and 1000 DN s$^{-1}$ (in blue). (f)-(i): observations from AIA 131~\AA. Panel (j): DEM weighted average temperature, showing the hot structure that appeared at 02:38~UT, during the filament interaction. Panels (k)-(m): formation of a bright front ahead of the erupting filaments, observed in AIA 171~\AA, see also Movie~3. (n): A partial halo CME associated with this eruption, viewed from SOHO/LASCO C2. The box regions in panels (m) and (n) indicate the field-of-view size of the corresponding panels.}
    \label{fig:fig3}
  \end{center}
\end{figure*}
 
\subsection{Configuration}
In order to reduce the impact of projection effects on the line-of-sight (LOS) magnetogram due to the location of AR~11777 near the solar limb (see the small panel in Figure~\ref{fig:fig1}(c)), we utilize a LOS magnetogram obtained four days later, when the projection effects were minimal, and differentially rotated to the same time as that of Figure~\ref{fig:fig1}(a) and displayed in Figure~\ref{fig:fig1}(c). The related PIL in this active region is determined based on the smoothed LOS magnetogram and indicated on the figure with a thick line. Two filaments were suspended above the same segment of the PIL: one filament was lower and shorter, and the other was higher and longer, as seen in Figure~\ref{fig:fig1}(a). These two filaments are the focus of the present study. In addition, a related thin loop-like structure (LLS) is observed to lie above the upper filament.

The observations during one day before the eruption reveal the formation of this complex configuration. The upper filament began to rise to a higher altitude after 11:00 UT on 2013 June 20. The lower filament started to ascend slowly about 15 hours later. During this interval, several strands of material were observed to transfer from the lower filament to the upper one, similar to previous studies on double-decker filaments (\citealp{liu2012slow,zhu2014eruption}). A height-time stack plot (Figure~\ref{fig:fig2}(b)) was generated from a series of SDO/AIA 193~\AA ~images along a slit marked in Figure~\ref{fig:fig2}(a). Each material transfer is indicated by an upward arrow. A few negative slopes, denoted by the downward arrows, correspond to mass drainage (see Movie~2). These observations suggest that both filaments were magnetically connected and associated with the same filament channel. 

The LLS was observed to be close to but separated from the upper filament shortly following a material transfer episode that occurred at around 23:40 UT on 2013 June 20, as seen in Movie~1, indicating that the LLS was closely related to the upper filament. After 23:50 UT, the upper filament became very faint, possibly due to mass drainage and/or heating, until it reappeared in the AIA 193~\AA ~and 304~\AA ~about two hours later (see Figure~\ref{fig:fig2}(b)).

 \begin{figure*}[ht!]
  \begin{center}
    \includegraphics[bb =  44 273 465 565, clip, height=0.7\textwidth]{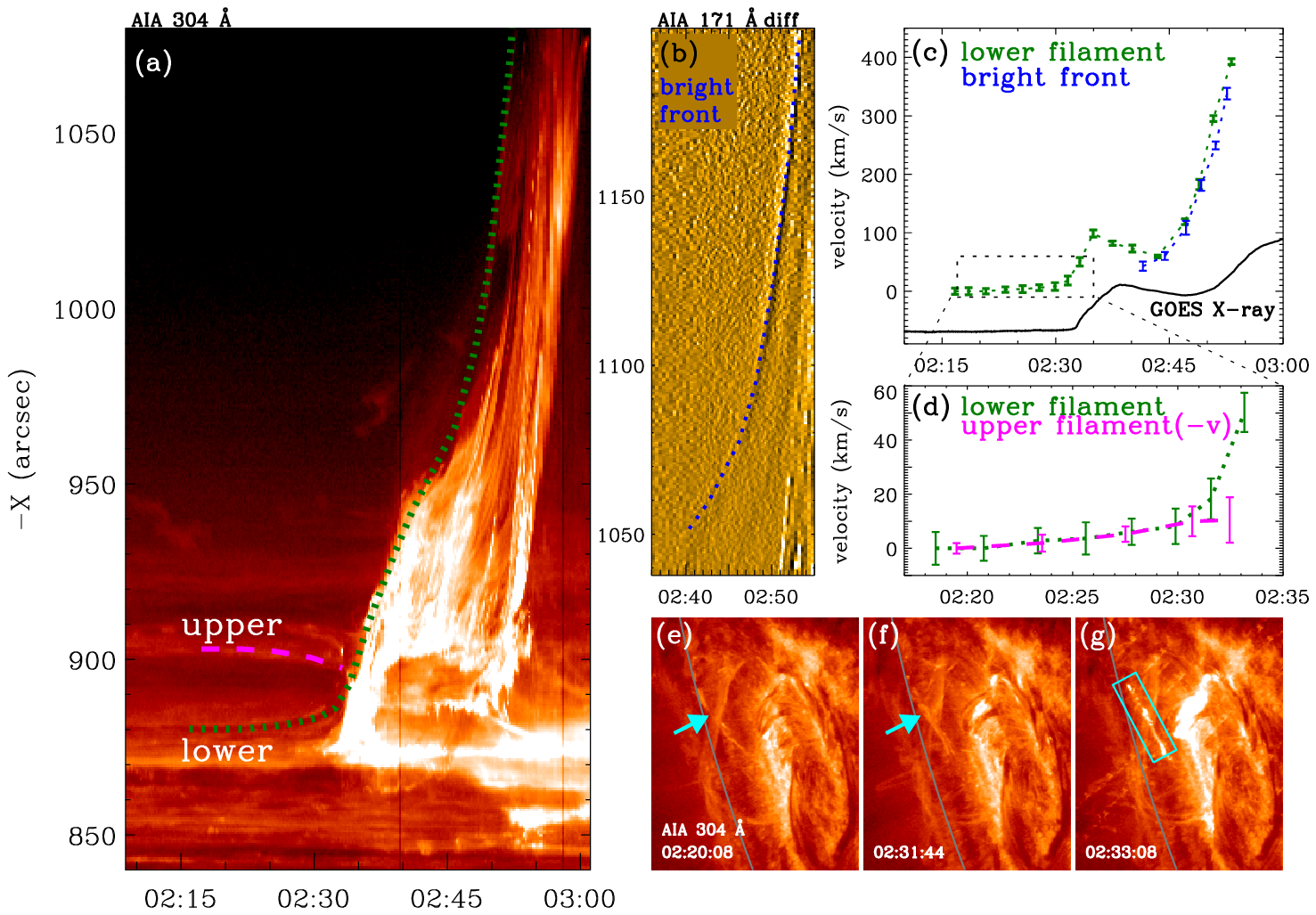}
    \caption[fig4]{Motions of the two filaments and a bright front ahead of the erupting filaments. (a): A stack plot along a cut denoted by the dash-dot line in Figure~\ref{fig:fig3}(b). The dotted line in (a) is used to denote the trajectory of the lower filament, and dashed line for the upper filament. (b): A stack plot of the bright front in Figure~\ref{fig:fig3}. (c): Velocities of the eruptive lower filament and the bright front. The normalized X-ray flux from GOES 1.6-12.4 keV is indicated by the black curve. A longer GOES X-ray profile can be found in Figure~\ref{fig:fig5}(a). (d): the velocities (absolute values) of both filaments before interaction, indicating their converging motion. (e)-(g): brightening of the eastern leg of the LLS during the interaction, displaying an apparent crossing over the upper filament to the foreground.}
    \label{fig:fig4}
  \end{center}
\end{figure*}

With the data from both SDO and STEREO-B (Figures~\ref{fig:fig1}(d) and (e)), the three-dimensional positions of both filaments before the eruption were measured by the SolarSoft routine \textsf{scc\_measure} (\citealp{thompson20123d}; \citealp{li2011three}). However, this analysis was not applied to the LLS or the two ends of the upper filament which were indistinguishable against the background emission when seen from STEREO-B. The reconstructed locations are rotated to the center of the solar disk and shown in Figure~\ref{fig:fig1}(f). When seen from above as in Figure~\ref{fig:fig1}(g), the upper filament appeared as a ``C'' shape. There was a crossing angle of around 20$^{\circ}$ between the projections of the two filaments onto the solar surface.

The lower filament began to rise at around 02:20 UT on 2013 June 21. The evolution of this filament eruption is displayed in Figure~\ref{fig:fig3}. A ``slit'' is placed along the East-West direction (dash-dot line in Figure~\ref{fig:fig3}(b)) in a series of AIA 304~\AA ~images to generate a stack plot as shown in Figure~\ref{fig:fig4}(a), in which the height of the lower filament is tracked by the dotted curve. From the time derivative of this height-time evolution, the velocity of the filament ascent is determined, and is found to exhibit four distinct phases (Figure~\ref{fig:fig4}(c)): 1) a slow-rise phase from 02:20-02:30 UT, with velocity below 10 km~s$^{-1}$, 2) a short fast-rise phase, from 02:30-02:35 UT, with velocity changing from $\sim$10 to 90 km~s$^{-1}$, 3) a minor relaxation phase with velocity decreasing from 90 to 60 km~s$^{-1}$, between 02:35-02:43 UT, and 4) another fast-rise phase starting from 02:43 UT, with the velocity increasing to $\sim$400 km~s$^{-1}$ by around 02:53 UT.  

A bright front ahead of the erupting filaments was observed to form at $\sim$02:38 UT, as seen in AIA 171~\AA ~(Figures~\ref{fig:fig3}(k)-(m) and Movie~3). With a stack plot (Figure~\ref{fig:fig4}(b)) along the same location in Figure~\ref{fig:fig3}(b), the velocity of the bright front is determined and shown in Figure~\ref{fig:fig4}(c). Its value increased from 40 km~s$^{-1}$ at 02:41 UT to 340 km~s$^{-1}$ at 02:53 UT. This velocity is slightly slower than that of the erupting filament, indicating that it is a compression front driven by the erupting filaments. The successful eruption was associated with a partial halo CME (Figure~\ref{fig:fig3}(n)) reported by Solar and Heliospheric Observatory\footnote{\url{http://cdaw.gsfc.nasa.gov/CME_list/daily_movies/2013/06/21/}}. A similar bright front was reported by \citet{zhang2012observation}, and was interpreted as the enveloping front of the CME associated with a filament eruption.

 \begin{figure*}[ht!]
  \begin{center}
      \includegraphics[bb = 46 156 656 661, clip, width=0.98\textwidth]{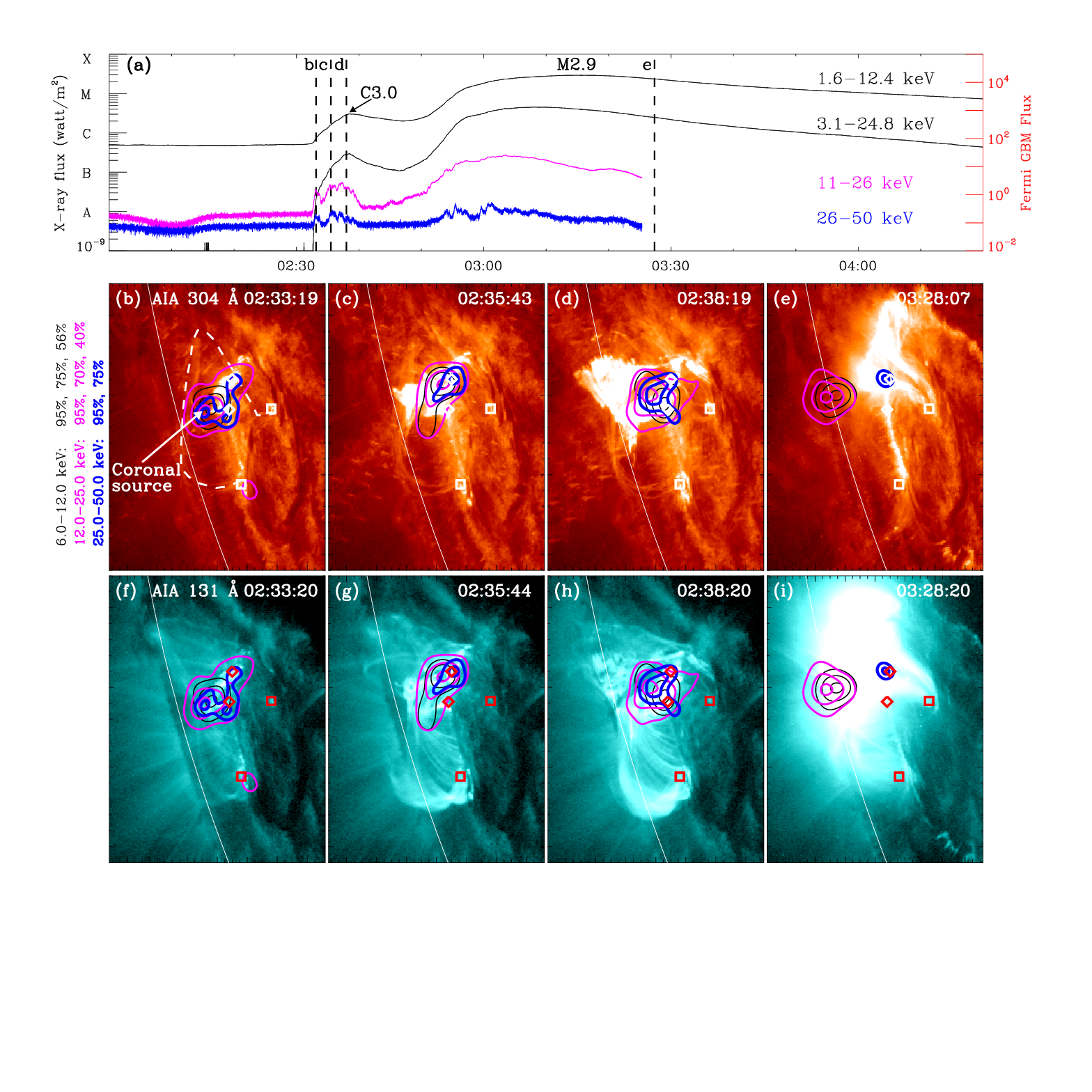}
    \caption[fig5]{Panel (a): Production of two consecutive solar flares. Lower panels: RHESSI X-ray observations overlaid on the EUV images from AIA 304~\AA ~and 131~\AA. The X-ray images were reconstructed using the CLEAN algorithm. The contours at 25-50 keV are marked with blue color, 12-25 keV with pink, 6-12 keV with black. The contour levels of the maximum value for their corresponding energy bins are denoted to the left of (b). The footpoints of the upper and the lower filaments are marked with symbols of squares and diamonds, respectively. The dashed line in (b) indicates the location of the upper filament at that time.}
    \label{fig:fig5}
  \end{center}
\end{figure*}

Solar flaring that was related to this filament eruption displayed two peaks in the GOES X-ray flux, as shown in Figure~\ref{fig:fig5}(a), a C3.0 peak followed by a transition to a second, much larger, M2.9 peak. As both of them show distinct peaks in soft and hard X-rays, and the hard X-ray lightcurves roughly returned to the background level between them (Figure~\ref{fig:fig5}(a)), we treat them as two distinct solar flares overlapping in time. In the remainder of this section, we investigate in detail the interaction between the two filaments and the evolution of their eruptions.

\subsection{Interaction Between the Two Filaments}

\textit{Plasma heating}. Prior to the eruption, the upper filament, which appeared in emission in AIA 304~\AA ~(Figure~\ref{fig:fig3}(a)), was barely identifiable in the hot EUV channels such as AIA 131~\AA ~(Figure~\ref{fig:fig3}(f)), possibly due to its lower temperature and density.  However at the time of the filament-filament interaction, $\sim$02:32~UT on June 21, a complex region of hot plasma appeared in the vicinity of the upper filament and the LLS, as seen in AIA 131~\AA ~(Figures~\ref{fig:fig3}(g)-(i)). The observation in the AIA 1600~\AA ~channel shows an extended chromospheric ribbon (Figure~\ref{fig:fig3}(e)), which is composed of the footpoints of numerous bright loops (Figures~\ref{fig:fig3}(g) and (h)) connected through this hot plasma region.  At the same time, all four filament footpoints (marked by diamonds and squares in Figure~\ref{fig:fig3}(e)) are found to be brightening in AIA 1600~\AA. The enhancement in the hot plasma was most evident at around 02:38 UT (Figure~\ref{fig:fig3}(h)). As AIA 131~\AA ~is sensitive to both hot and cold plasmas at 10 MK and 0.4 MK respectively, we reconstructed the differential emission measure (DEM) of the region using the code developed by \citet{plowman2013fast} and calculated the DEM-weighted average temperature $\langle$T$\rangle$ (\eg \citealp{cheng2012differential,guidoni2015temperature,Gou2015Do}), as shown in Figure~\ref{fig:fig3}(j). The value of $\langle$T$\rangle$ in the vicinity of the upper filament ranges from 7 to 12 MK.  

\textit{Converging motion}. The motions of the two filaments are determined by tracking the moving features evident in the generated stack plot (Figure~\ref{fig:fig4}(a)), with derived velocities shown in Figure~\ref{fig:fig4}(d). Immediately prior to the filament interaction at 02:32 UT, the descending upper filament structure reached a speed of 10 km~s$^{-1}$. At this time, the lower filament was rising at $\sim$40 km~s$^{-1}$. This observation indicates that the filaments converged with a relatively rapid rise of the lower filament and a slow descent of the upper filament. 

It is noticeable that the eastern leg of the LLS was brightening in AIA 304~\AA ~as the interaction was in progress (Figures~\ref{fig:fig4}(e)-(g)). This leg, lying behind the upper filament before the interaction, appeared to be crossing through the upper filament to the foreground after $\sim$02:32 UT (Figure~\ref{fig:fig4}(f) and movie~3). The brightening and the apparent crossing of the LLS indicate its direct involvement in the interaction with the rising lower filament.

\textit{The GOES C3.0 flare}. Coincident with the filament interaction, the GOES satellite detected the onset of a C3.0 flare. A hard X-ray burst observed by the Fermi
Gamma-ray Burst Monitor (GBM; \citealp{meegan2009fermi}) initiated at 02:32 UT and lasted for about 7 minutes (Figure~\ref{fig:fig5}(a)). Three intervals were chosen to study the spatial evolution of the RHESSI X-ray sources (denoted by the first three dashed lines in Figure~\ref{fig:fig5}(a), and corresponding to the early, middle and late intervals of the burst, respectively). The results are displayed in Figures~\ref{fig:fig5}(b)-(d) and (f)-(h).
At 02:33 UT (Figures~\ref{fig:fig5}(b) and (f)), a coronal X-ray source was detected near the interface of the two filaments,  emitting in both soft and hard X-rays. Three footpoint sources are identified: two of them were located near the footpoints of the lower filament (indicated by two diamond symbols), and a third one (marked by the southern square symbol), which was relatively weak, lay near the southern footpoint of the upper filament. Beginning at 02:34 UT, the northmost footpoint source became dominant in the hard X-ray emission (Figures~\ref{fig:fig5}(c) and (g)), indicating that this flare is asymmetric (\citealp{alexander2006temporal,coyner2009implications}).  At 02:38 UT (Figures~\ref{fig:fig5}(d) and (h)), with the decay of the hard X-ray flux (Figure~\ref{fig:fig5}(a)), the nature of the X-ray sources are difficult to determine due to the complexity of the configuration and possible line-of-sight confusion of the various features.

\subsection{M2.9 Flare and Field Line Shrinkage}
An M2.9 solar flare began at around 02:45 UT, five minutes after the decay phase of the C3.0 flare (Figure~\ref{fig:fig5}(a)), and lasted for more than 3 hours till $\sim$06:00 UT. RHESSI observations taken near this flare peak are shown in Figures~\ref{fig:fig5}(e) and (i). A loop-top source was located above the brightening arcade, with a projection of a hard X-ray source near the northern footpoint of the lower filament.

 \begin{figure*}[ht!]
  \begin{center}
      \includegraphics[bb =  31 46 537 497, clip, height=0.7\textwidth]
{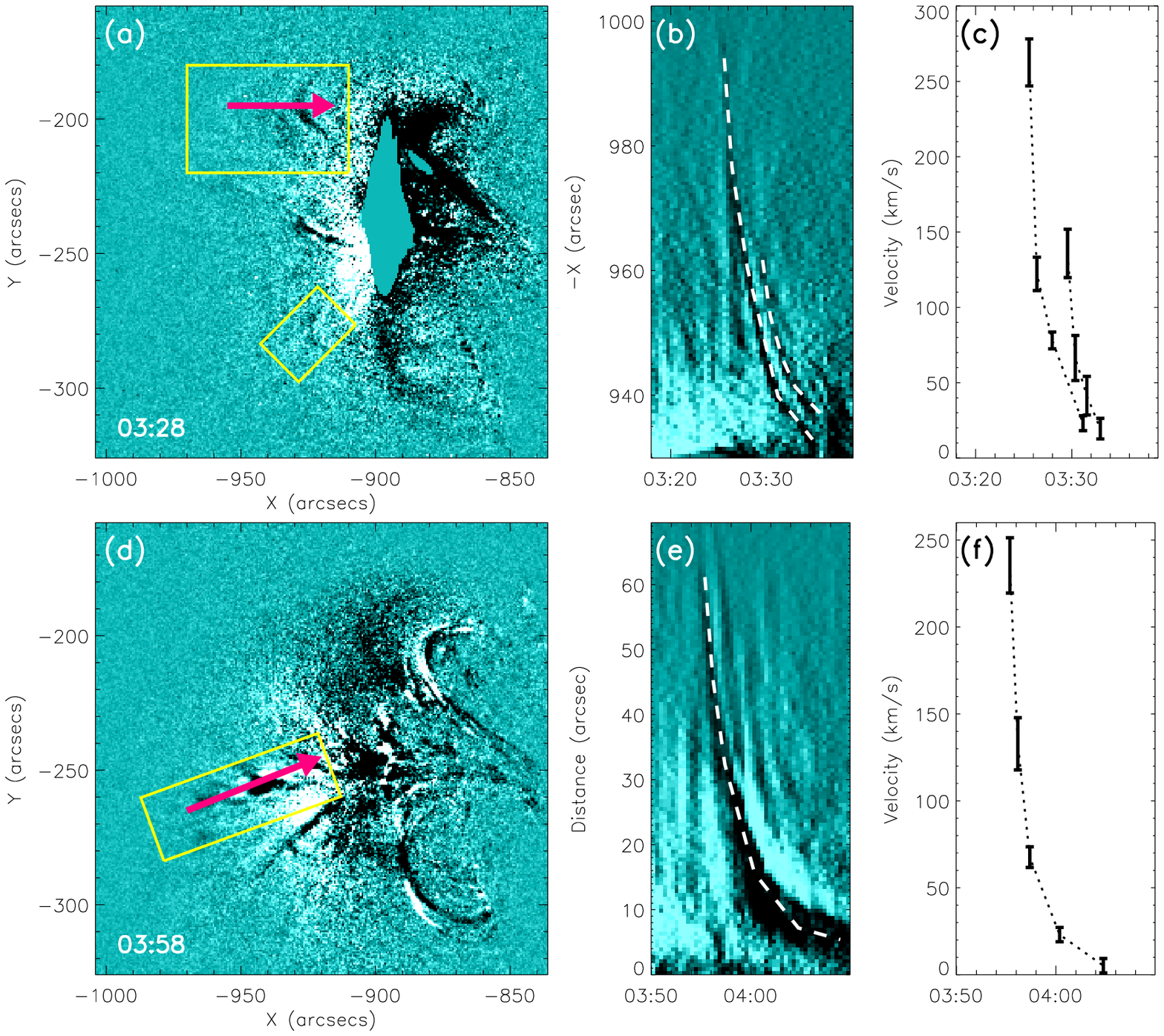} 
  \end{center}    
    \caption[fig6]{Observations of the magnetic loop shrinkage (upper panels) and descending dark voids (lower panels) in the M2.9 flare. (a): running difference of AIA 131~\AA ~images. Two locations with loop shrinkage are identified, delineated in two boxes. A cut along the arrow in the top box is chosen to generate a height-time stack plot shown in (b). The direction of the arrow indicates that the loop was relaxing towards the post-flare arcade. (b): Trajectories of two retracting loops, indicated by two dashed lines. The derived velocities are shown in (c). Lower panels: observation of a descending dark void. Same as in upper panels. Movie~4 clearly shows these dynamic features.}
    \label{fig:fig6}
\end{figure*}

\textit{Field Line shrinkage} (or \textit{loop retraction}, \eg \citealp{forbes1996reconnection,mckenzie1999x}) was observed during this M flare. Specifically, two  locations are identified at the northern and southern ends of the post-flare arcade at 03:28 UT (Figure~\ref{fig:fig6}(a) and Movie~4). In both regions,  shrinking loops are observed to retract towards the post-flare loops.  
The velocity of the retracting loops is shown via a stack plot from a cut along the East-West direction (along the arrow in Figure~\ref{fig:fig6}(a), with the arrow indicating the direction of motion). Two moving features, denoted by the dashed lines in Figure~\ref{fig:fig6}(b), correspond to two distinct retracting loops. The velocities of both loops were observed to be decreasing as they approached the post-flare loops, changing from a few hundreds km~s$^{-1}$ to rest (Figure~\ref{fig:fig6}(c)).

It is also interesting to note that several tadpole-like dark voids were evident in AIA 131~\AA ~(Movie~4),  travelling downward successively, in this flare. These tadpole-like voids are also called supra-arcade downflows (SADs; \eg \citealp{mckenzie1999x,liu2013dynamical}).  The box region in Figure~\ref{fig:fig6}(d) shows one example of these SADs. The location of the SADs, directly above 
the center of the flare loop arcade, was distinct from the region of the observed pronounced field line shrinkage. The trajectory of the SAD is determined, again, from the stack plot method (Figure~\ref{fig:fig6}(e)) and the velocity derived is shown in Figure~\ref{fig:fig6}(f). The SAD structure descended with a initial velocity of 240 km~s$^{-1}$ at 03:57 UT until it came to rest at the flare loop region ten minutes later. Additional SADs were still observable by the end of this flare, at $\sim$06:00 UT.

\section{\uppercase{Interpretation and Discussion}}

\subsection{Filament-Filament Interaction}

\citet{kumar2010evolution} reported a converging motion of two filaments at a speed of $\sim$10 km~s$^{-1}$ during their slingshot reconnection. This value is typical for reconnection inflows detected at EUV wavelengths \citep{yokoyama2001clear,liu2010reconnecting}, and is consistent with the theoretical expectations of \citet{petschek1964physics}. In the present study, immediately prior to the direct interaction of the two filaments, the upper filament was observed to  descend at $\sim$10 km~s$^{-1}$, while the eruptive lower filament rose up with  a velocity of $\sim$40 km~s$^{-1}$. The velocities of the observed converging motions are comparable with previous studies. 

While the dynamical motions are similar to those observed in previous observations, it would appear that in this case the driving mechanisms for the converging motion are different. In the study of \citet{kumar2010evolution}, the approaching filaments are thought to be driven by slow photospheric motions. The filament-filament interaction reported by \citet{Jiang2013} is initiated by one of the filaments that erupted. In the event  reported here, the lower filament is eruptive, similar to \citet{Jiang2013}, but the descending motion of the upper filament requires different explanations. A possible candidate is the $\textbf{J}\times\textbf{B}$ force  between the two filament current systems. This scenario is expected from the simulation of a full eruption of an unstable double flux rope (see Figure~3 in \citealp{kliem2014slow}) and the laboratory experiment of two parallel current channels by \citet{intrator2009experimental}. An alternative explanation calls for an increase in the magnetic tension of the strapping fields which regulate the height of the upper filament, though this scenario lacks observational evidence as the evolution of the strapping field is difficult to diagnose in this event. Furthermore, the eastern leg of the LLS, which appeared to pass through the upper filament during the interaction, might serve to  depress the upper filament and contribute to its observed decent. To our knowledge, this is the first time that a converging motion has been observed between filaments in a double-decker configuration.

The newly formed hot region of plasma in the vicinity of the upper filament consists of numerous hot loops (Figure~\ref{fig:fig3}(h)).  This hot plasma may be heated in a quasi-separatrix layer (\citealp{Demoulin1996}), which wraps around the upper filament, separating its twisted magnetic fields from the outer, untwisted fields. As the two filaments interact, accelerated electrons stream along the loops in this layer and deposit energy into the footpoints of these loops  as the electrons are stopped by the dense chromosphere, producing the extended ribbon observed in AIA 1600~\AA ~(Figure~\ref{fig:fig3}(e)).

A coronal hard X-ray source is identifiable near the interaction interface of the filaments, as seen in Figure~\ref{fig:fig5}(b). Due to the limited resolution of RHESSI, this coronal source could have two origins: ongoing reconnection between the interacting fields of the two filaments, and/or a loop-top source located at or above the apex of the flare loops (\eg \citealp{Masuda1994,liu2013plasmoid}). However, the coronal source  dominated the hard X-ray emission at the onset of this C flare, unlike a typical loop-top source which is usually fainter than its associated footpoint sources (\citealp{petrosian2002loop}). Thus, we suggest that the observed coronal source at least partially resulted from ongoing reconnection between the magnetic fields around both filaments. It is worth noting that the coronal hard X-ray source is barely detected by RHESSI in the later phase of the C3.0 flare (Figures~\ref{fig:fig5}(c) and (d)). This may be because the footpoint source had grown in intensity to dominate over the coronal source and the limited RHESSI dynamic range would make it extremely difficult to detect any week coronal emission (\eg \citealp{sui2004evidence}). Thus, it is difficult to determine whether the two filaments were still interacting or not in the later stage of the eruption.

Both the upper and lower filaments are suspended above the same region of AR~11777 and oriented with a small projected angle of around 20 degrees between them. In the simulations of \citet{Linton2001} and \citet{kliem2014slow}, such a configuration would result in the merger of the two filaments following the initial interaction. In our study, there are no evident signatures for the other types of interaction  discussed by \citet{Linton2001}  (\ie bounce, tunnel and slingshot). As the lower eruptive filament interacts with the upper, the combined system continues to rise  until both finally erupt away. From this point of view, both filaments are assumed to have merged, at least partially, to form a complicated eruptive structure with four distinct footpoints rooted in the chromosphere.

\subsection{Field Line Shrinkage in the M Flare}
Both the observed field line shrinkage and the SADs are generally thought to be consequences of the newly reconnected evacuating flux tubes retracting from the reconnection site above the bright arcade of loops during solar flares (\citealp{vsvestka1987multi,forbes1996reconnection,mckenzie1999x,mckenzie2000supra}). Whether the retractions appear as loop shrinkage or sunward dark voids, depends on the different viewpoints (Figure~2 in  \citealp{savage2011quantitative}): when the retracting flux tubes are viewed along the axis of the flare arcade (\ie face-on), they appear as shrinkage; when viewed perpendicular to the axis (\ie side-on), they would be observed as dark voids (\citealp{savage2011quantitative,savage2012re}). \citet{warren2011observations} demonstrated this speculation using STEREO and SDO observations. In our study, the post-flare loop arcade appears C-shaped (see Figure~\ref{fig:fig6}(d)), and fans out towards the dispersed positive polarities in the east, with the footpoints of individual loops concentrated near the compact sunspot in the west. Due to the particular shape of this arcade, our view towards the loops changes from face-on in the south to side-on at the center, and becomes face-on again in the north of the arcade.  As a result, field line shrinkages at the two ends of the arcade were observed in a face-on view, and SADs in the center in a side-on view. \citet{warren2011observations} also reported a similar result in a $\Gamma$-shaped post-flare loop arcade, though the configuration here may be  more complicated. The initial velocities of the two loop shrinkage and one SAD events are 260, 120 and 240 km~s$^{-1}$, respectively. These values are typical of SADs (\eg \citealp{savage2011quantitative}).

The observation of loop shrinkage and SADs indicates that a large-scale vertical current sheet existed above the flare loop arcade (\citealp{mckenzie1999x}; \citealp{liu2013plasmoid}; \citealp{liu2013dynamical}). In our study, the overlying magnetic field lines might become highly stretched due to the filament eruption, forming a vertical current sheet underneath it. This scenario is expected in the standard flare model (\citealp{kopp1976magnetic}). Thus, we suggest that the process for this M2.9 flare follows the standard model.

\section{\uppercase{Conclusion}}
We reported the evolution of a filament eruption on 2013 June 21, which involved a few rare features, including the interaction between two filaments with a double-decker configuration, the simultaneous appearance of an apparent crossing of a loop-like structure though the upper filament, and the formation of a bright front driven by this eruption. Based on the observations of the converging motions of the two filaments, and the subsequent appearance of a hot plasma layer and a coronal hard X-ray source near the interaction interface, we suggest that the magnetic fields associated with the two filaments reconnected with each other during their interaction.

The interaction between the two filaments with a small contact angle of around 20 degrees supports the merger scenario, being consistent with theoretical studies by \citet{Linton2001} and \cite{kliem2014slow}. How complete (fully or partial) the merger of the two filaments was and how long the interaction lasted remain unclear and needs to be investigated in future studies.

The complex structure that was formed by the merger of the two filaments subsequently erupted away generating a large M2.9 flare. The resulting  post-flare loop arcade was C-shaped so that we were able to observe simultaneous loop shrinkage and SADs at different parts of the structure.  The observation of loop shrinkage and SADs indicates that magnetic reconnection occurs above the post-flare loop arcade, consistent with the standard flare model.

\ \\
\ \\

\acknowledgments
The authors would like to thank the referee for many valuable comments that helped improve this paper. RTJM and CZ are funded by NSF-CAREER 1255024. DA and CZ acknowledges support by the NSF SHINE grant AGS-1061899. RL acknowledges the Thousand Young Talents Programme of China, NSFC 41222031, NSFC 41474151 and NSF AGS-1153226.  We also thank SDO, STEREO, SOHO and RHESSI for the data support.

\bibliographystyle{yahapj} 
\bibliography{rc2bib}		

\begin{thebibliography}{53}
\providecommand\natexlab[1]{#1}
\providecommand\JournalTitle[1]{#1}

\bibitem[{Alexander \& Coyner(2006)}]{alexander2006temporal}
Alexander, D., \& Coyner, A.~J. 2006,
  \href{http://adsabs.harvard.edu/abs/2006ApJ...640..505A}{\JournalTitle{\apj},
  640, 505}

\bibitem[{Alexander {et~al.}(2006)Alexander, Liu, \&
  Gilbert}]{alexander2006hard}
Alexander, D., Liu, R., \& Gilbert, H.~R. 2006,
  \href{http://dx.doi.org/10.1086/508137}{\JournalTitle{\apj}, 653, 719}

\bibitem[{Bone {et~al.}(2009)Bone, van Driel-Gesztelyi, Culhane, Aulanier, \&
  Liewer}]{bone2009formation}
Bone, L., van Driel-Gesztelyi, L., Culhane, J., Aulanier, G., \& Liewer, P.
  2009,
  \href{http://dx.doi.org/10.1007/s11207-009-9427-5}{\JournalTitle{\solphys},
  259, 31}

\bibitem[{Chandra {et~al.}(2011)Chandra, Schmieder, Mandrini, D{\'e}moulin,
  Pariat, T{\"o}r{\"o}k, \& Uddin}]{chandra2011homologous}
Chandra, R., Schmieder, B., Mandrini, C., {et~al.} 2011,
  \href{http://link.springer.com/article/10.1007/s11207-010-9670-9}{\JournalTitle{\solphys},
  269, 83}

\bibitem[{Cheng {et~al.}(2012)Cheng, Zhang, Saar, \&
  Ding}]{cheng2012differential}
Cheng, X., Zhang, J., Saar, S., \& Ding, M. 2012,
  \href{http://dx.doi.org/10.1088/0004-637X/761/1/62}{\JournalTitle{\apj}, 761,
  62}

\bibitem[{Coyner \& Alexander(2009)}]{coyner2009implications}
Coyner, A.~J., \& Alexander, D. 2009,
  \href{http://adsabs.harvard.edu/abs/2009ApJ...705..554C}{\JournalTitle{\apj},
  705, 554}

\bibitem[{D{\'e}moulin {et~al.}(1996)D{\'e}moulin, Priest, \&
  Lonie}]{Demoulin1996}
D{\'e}moulin, P., Priest, E.~R., \& Lonie, D.~P. 1996,
  \href{http://dx.doi.org/10.1029/95ja03558}{\JournalTitle{\jgr}, 101, 7631}

\bibitem[{DeVore {et~al.}(2005)DeVore, Antiochos, \&
  Aulanier}]{devore2005solar}
DeVore, C.~R., Antiochos, S.~K., \& Aulanier, G. 2005,
  \href{http://adsabs.harvard.edu/abs/2005ApJ...629.1122D}{\JournalTitle{\apj},
  629, 1122}

\bibitem[{Forbes(2000)}]{forbes2000review}
Forbes, T. 2000,
  \href{http://adsabs.harvard.edu/abs/2000JGR...10523153F}{\JournalTitle{\jgr},
  105, 23153}

\bibitem[{Forbes \& Acton(1996)}]{forbes1996reconnection}
Forbes, T., \& Acton, L. 1996,
  \href{http://dx.doi.org/10.1086/176896}{\JournalTitle{\apj}, 459, 330}

\bibitem[{Gaizauskas(1998)}]{gaizauskas1998filament}
Gaizauskas, V. 1998,
  \href{http://adsabs.harvard.edu/abs/1998ASPC..150..257G}{in IAU Colloq. 167:
  New Perspectives on Solar Prominences, Vol. 150}, 257

\bibitem[{Gaizauskas {et~al.}(1997)Gaizauskas, Zirker, Sweetland, \&
  Kovacs}]{gaizauskas1997formation}
Gaizauskas, V., Zirker, J., Sweetland, C., \& Kovacs, A. 1997,
  \href{http://adsabs.harvard.edu/abs/1997ApJ...479..448G}{\JournalTitle{\apj},
  479, 448}

\bibitem[{Gou {et~al.}(2015)Gou, Liu, \& Wang}]{Gou2015Do}
Gou, T., Liu, R., \& Wang, Y. 2015,
  \href{http://dx.doi.org/10.1007/s11207-015-0750-8}{\JournalTitle{\solphys}}

\bibitem[{Guidoni {et~al.}(2015)Guidoni, McKenzie, Longcope, Plowman, \&
  Yoshimura}]{guidoni2015temperature}
Guidoni, S., McKenzie, D., Longcope, D., Plowman, J., \& Yoshimura, K. 2015,
  \href{http://dx.doi.org/10.1088/0004-637X/800/1/54}{\JournalTitle{\apj}, 800,
  54}

\bibitem[{Hoeksema {et~al.}(2014)Hoeksema, Liu, Hayashi, Sun, Schou, Couvidat,
  Norton, Bobra, Centeno, Leka, {et~al.}}]{hoeksema2014helioseismic}
Hoeksema, J.~T., Liu, Y., Hayashi, K., {et~al.} 2014,
  \href{http://adsabs.harvard.edu/abs/2014SoPh..tmp...57H}{\JournalTitle{\solphys},
  289, 3483}

\bibitem[{Intrator {et~al.}(2009)Intrator, Sun, Lapenta, Dorf, \&
  Furno}]{intrator2009experimental}
Intrator, T., Sun, X., Lapenta, G., Dorf, L., \& Furno, I. 2009,
  \href{http://dx.doi.org/10.1038/nphys1300}{\JournalTitle{Nature Phys.}, 5,
  521}

\bibitem[{Jiang {et~al.}(2013)Jiang, Hong, Yang, Bi, Zheng, Yang, Li, \&
  Yang}]{Jiang2013}
Jiang, Y., Hong, J., Yang, J., {et~al.} 2013,
  \href{http://dx.doi.org/10.1088/0004-637x/764/1/68}{\JournalTitle{\apj}, 764}

\bibitem[{Jiang {et~al.}(2014)Jiang, Yang, Wang, Ji, Liu, Li, \&
  Li}]{jiang2014interaction}
Jiang, Y., Yang, J., Wang, H., {et~al.} 2014,
  \href{http://dx.doi.org/10.1088/0004-637X/793/1/14}{\JournalTitle{\apj}, 793,
  14}

\bibitem[{Kliem {et~al.}(2014)Kliem, T{\"o}r{\"o}k, Titov, Lionello, Linker,
  Liu, Liu, \& Wang}]{kliem2014slow}
Kliem, B., T{\"o}r{\"o}k, T., Titov, V.~S., {et~al.} 2014,
  \href{http://iopscience.iop.org/0004-637X/792/2/107}{\JournalTitle{\apj},
  792, 107}

\bibitem[{Kopp \& Pneuman(1976)}]{kopp1976magnetic}
Kopp, R., \& Pneuman, G. 1976,
  \href{http://dx.doi.org/10.1007/BF00206193}{\JournalTitle{\solphys}, 50, 85}

\bibitem[{Kumar {et~al.}(2010)Kumar, Manoharan, \& Uddin}]{kumar2010evolution}
Kumar, P., Manoharan, P., \& Uddin, W. 2010,
  \href{http://iopscience.iop.org/0004-637X/710/2/1195}{\JournalTitle{\apj},
  710, 1195}

\bibitem[{Lemen {et~al.}(2012)Lemen, Title, Akin, Boerner, Chou, Drake, Duncan,
  Edwards, Friedlaender, Heyman, {et~al.}}]{Lemen2012}
Lemen, J.~R., Title, A.~M., Akin, D.~J., {et~al.} 2012,
  \href{http://adsabs.harvard.edu/abs/2012SoPh..275...17L}{\JournalTitle{\solphys},
  275, 17}

\bibitem[{Li {et~al.}(2011)Li, Zhang, Zhang, \& Yang}]{li2011three}
Li, T., Zhang, J., Zhang, Y., \& Yang, S. 2011,
  \href{http://dx.doi.org/10.1088/0004-637X/739/1/43}{\JournalTitle{\apj}, 739,
  43}

\bibitem[{Linton(2006)}]{Linton2006}
Linton, M.~G. 2006,
  \href{http://dx.doi.org/10.1029/2006ja011891}{\JournalTitle{\jgr}, 111}

\bibitem[{Linton {et~al.}(2001)Linton, Dahlburg, \& Antiochos}]{Linton2001}
Linton, M.~G., Dahlburg, R.~B., \& Antiochos, S.~K. 2001,
  \href{http://dx.doi.org/10.1086/320974}{\JournalTitle{\apj}, 553, 905}

\bibitem[{Litvinenko \& Martin(2000)}]{litvinenko2000magnetic}
Litvinenko, Y.~E., \& Martin, S.~F. 2000,
  \href{http://dx.doi.org/10.1023/A:1005284116353}{in Physics of the Solar
  Corona and Transition Region} (Springer), 45

\bibitem[{Liu(2013)}]{liu2013dynamical}
Liu, R. 2013,
  \href{http://adsabs.harvard.edu/abs/2013MNRAS.434.1309L}{\JournalTitle{\mnras},
  434, 1309}

\bibitem[{Liu {et~al.}(2012)Liu, Kliem, T{\"o}r{\"o}k, Liu, Titov, Lionello,
  Linker, \& Wang}]{liu2012slow}
Liu, R., Kliem, B., T{\"o}r{\"o}k, T., {et~al.} 2012,
  \href{http://iopscience.iop.org/0004-637X/756/1/59}{\JournalTitle{\apj}, 756,
  59}

\bibitem[{Liu {et~al.}(2010)Liu, Lee, Wang, Stenborg, Liu, \&
  Wang}]{liu2010reconnecting}
Liu, R., Lee, J., Wang, T., {et~al.} 2010,
  \href{http://adsabs.harvard.edu/abs/2010arXiv1009.4912L}{\JournalTitle{\apjl},
  723, L28}

\bibitem[{Liu {et~al.}(2013)Liu, Chen, \& Petrosian}]{liu2013plasmoid}
Liu, W., Chen, Q., \& Petrosian, V. 2013,
  \href{http://dx.doi.org/10.1088/0004-637X/767/2/168}{\JournalTitle{\apj},
  767, 168}

\bibitem[{Martens \& Zwaan(2001)}]{martens2001origin}
Martens, P.~C., \& Zwaan, C. 2001,
  \href{http://adsabs.harvard.edu/abs/2001ApJ...558..872M}{\JournalTitle{\apj},
  558, 872}

\bibitem[{Masuda {et~al.}(1994)Masuda, Kosugi, Hara, \& Ogawara}]{Masuda1994}
Masuda, S., Kosugi, T., Hara, H., \& Ogawara, Y. 1994,
  \href{http://dx.doi.org/10.1038/371495a0}{\JournalTitle{Nature}, 371, 495}

\bibitem[{McKenzie(2000)}]{mckenzie2000supra}
McKenzie, D. 2000,
  \href{http://dx.doi.org/10.1023/A:1005220604894}{\JournalTitle{\solphys},
  195, 381}

\bibitem[{McKenzie \& Hudson(1999)}]{mckenzie1999x}
McKenzie, D., \& Hudson, H. 1999,
  \href{http://adsabs.harvard.edu/abs/1999ApJ...519L..93M}{\JournalTitle{\apjl},
  519, L93}

\bibitem[{Meegan {et~al.}(2009)Meegan, Lichti, Bhat, Bissaldi, Briggs,
  Connaughton, Diehl, Fishman, Greiner, Hoover, {et~al.}}]{meegan2009fermi}
Meegan, C., Lichti, G., Bhat, P., {et~al.} 2009,
  \href{http://dx.doi.org/10.1088/0004-637X/702/1/791}{\JournalTitle{\apj},
  702, 791}

\bibitem[{Petrosian {et~al.}(2002)Petrosian, Donaghy, \&
  McTiernan}]{petrosian2002loop}
Petrosian, V., Donaghy, T.~Q., \& McTiernan, J.~M. 2002,
  \href{http://dx.doi.org/10.1086/339240}{\JournalTitle{\apj}, 569, 459}

\bibitem[{Petschek(1964)}]{petschek1964physics}
Petschek, H. 1964, in Proc. AAS-NASA Symp., ed. Hess W. N. (Greenbelt,
  MD:NASA), Vol. 425

\bibitem[{Plowman {et~al.}(2013)Plowman, Kankelborg, \&
  Martens}]{plowman2013fast}
Plowman, J., Kankelborg, C., \& Martens, P. 2013,
  \href{http://dx.doi.org/10.1088/0004-637X/771/1/2}{\JournalTitle{\apj}, 771,
  2}

\bibitem[{Savage \& McKenzie(2011)}]{savage2011quantitative}
Savage, S.~L., \& McKenzie, D.~E. 2011,
  \href{http://adsabs.harvard.edu/abs/2011arXiv1101.1540S}{\JournalTitle{\apj},
  730, 98}

\bibitem[{Savage {et~al.}(2012)Savage, McKenzie, \& Reeves}]{savage2012re}
Savage, S.~L., McKenzie, D.~E., \& Reeves, K.~K. 2012,
  \href{http://dx.doi.org/10.1088/2041-8205/747/2/L40}{\JournalTitle{\apjl},
  747, L40}

\bibitem[{Schmieder {et~al.}(2004)Schmieder, Mein, Deng, Dumitrache, Malherbe,
  Staiger, \& Deluca}]{schmieder2004magnetic}
Schmieder, B., Mein, N., Deng, Y., {et~al.} 2004,
  \href{http://dx.doi.org/10.1007/s11207-004-1107-x}{\JournalTitle{\solphys},
  223, 119}

\bibitem[{Schou {et~al.}(2012)Schou, Scherrer, Bush, Wachter, Couvidat,
  Rabello-Soares, Bogart, Hoeksema, Liu, Duvall, {et~al.}}]{Schou2012}
Schou, J., Scherrer, P., Bush, R., {et~al.} 2012,
  \href{http://dx.doi.org/10.1007/978-1-4614-3673-7_11}{\JournalTitle{\solphys},
  275, 229}

\bibitem[{Su {et~al.}(2007)Su, Liu, Kurokawa, Mao, Yang, Zhang, \&
  Wang}]{su2007observation}
Su, J., Liu, Y., Kurokawa, H., {et~al.} 2007,
  \href{http://dx.doi.org/10.1007/s11207-007-0213-y}{\JournalTitle{\solphys},
  242, 53}

\bibitem[{Sui {et~al.}(2004)Sui, Holman, \& Dennis}]{sui2004evidence}
Sui, L., Holman, G.~D., \& Dennis, B.~R. 2004,
  \href{http://dx.doi.org/10.1086/422515}{\JournalTitle{\apj}, 612, 546}

\bibitem[{{\v{S}}vestka {et~al.}(1987){\v{S}}vestka, Fontenla, Machado, Martin,
  Neidig, \& Poletto}]{vsvestka1987multi}
{\v{S}}vestka, Z.~F., Fontenla, J.~M., Machado, M.~E., {et~al.} 1987,
  \href{http://dx.doi.org/10.1007/BF00214164}{\JournalTitle{\solphys}, 108,
  237}

\bibitem[{Thompson {et~al.}(2012)Thompson, Kliem, \&
  T{\"o}r{\"o}k}]{thompson20123d}
Thompson, W., Kliem, B., \& T{\"o}r{\"o}k, T. 2012,
  \href{http://dx.doi.org/10.1007/s11207-011-9868-5}{\JournalTitle{\solphys},
  276, 241}

\bibitem[{T\"or\"ok {et~al.}(2011)T\"or\"ok, Chandra, Pariat, Demoulin,
  Schmieder, Aulanier, Linton, \& Mandrini}]{Toeroek2011}
T\"or\"ok, T., Chandra, R., Pariat, E., {et~al.} 2011,
  \href{http://dx.doi.org/10.1088/0004-637x/728/1/65}{\JournalTitle{\apj}, 728}

\bibitem[{Wang \& Muglach(2007)}]{wang2007formation}
Wang, Y.-M., \& Muglach, K. 2007,
  \href{http://dx.doi.org/10.1086/520623}{\JournalTitle{\apj}, 666, 1284}

\bibitem[{Warren {et~al.}(2011)Warren, O'Brien, \&
  Sheeley~Jr}]{warren2011observations}
Warren, H.~P., O'Brien, C.~M., \& Sheeley~Jr, N.~R. 2011,
  \href{http://dx.doi.org/10.1088/0004-637X/742/2/92}{\JournalTitle{\apj}, 742,
  92}

\bibitem[{Wuelser {et~al.}(2004)Wuelser, Lemen, Tarbell, Wolfson, Cannon,
  Carpenter, Duncan, Gradwohl, Meyer, Moore, {et~al.}}]{wuelser2004euvi}
Wuelser, J.~P., Lemen, J.~R., Tarbell, T.~D., {et~al.} 2004,
  \href{http://dx.doi.org/10.1117/12.506877}{in Proc. SPIE, Vol. 5171}, 111

\bibitem[{Yokoyama {et~al.}(2001)Yokoyama, Akita, Morimoto, Inoue, \&
  Newmark}]{yokoyama2001clear}
Yokoyama, T., Akita, K., Morimoto, T., Inoue, K., \& Newmark, J. 2001,
  \href{http://dx.doi.org/10.1086/318053}{\JournalTitle{\apjl}, 546, L69}

\bibitem[{Zhang {et~al.}(2012)Zhang, Cheng, {et~al.}}]{zhang2012observation}
Zhang, J., Cheng, X., {et~al.} 2012,
  \href{http://dx.doi.org/10.1038/ncomms1753}{\JournalTitle{Nat. Commun.}, 3,
  747}

\bibitem[{Zhu \& Alexander(2014)}]{zhu2014eruption}
Zhu, C., \& Alexander, D. 2014,
  \href{http://dx.doi.org/10.1007/s11207-013-0349-x}{\JournalTitle{\solphys},
  289, 279}

\end{thebibliography}

\end{document}